# Oscillations in Quantum Entanglement During Rescattering

**Attila Czirják, Szilárd Majorosi, Judit Kovács, and Mihály G. Benedict**
*Department of Theoretical Physics, University of Szeged, H-6720 Szeged, Tisza Lajos krt. 84-86, Hungary*
*czirjak@physx.u-szeged.hu, benedict@physx.u-szeged.hu*

**Abstract:** We study the time evolution of quantum entanglement between an electron and its parent ion during the rescattering due to a strong few-cycle laser pulse. Based on a simple one-dimensional model, we compute the Neumann entropy during the process for several values of the carrier-envelope phase. The local maxima of the oscillations in the Neumann entropy coincide with the zero crossings of the electric field of the laser pulse. We employ the Wigner function to qualitatively explain the quantum dynamics of rescattering in the phase space.

## 1. Introduction

The collision of two particles is one of the simplest scenarios that create quantum entanglement [1-3]. This topic has growing relevance in the context of attosecond physics [4], since measurements performed on the scattered particles (e.g. electrons, ions or ionized molecules) [5-9] can reveal quantum features of the underlying processes [10-12].

In this contribution, we study the time dependence of quantum entanglement in a fundamental process of attosecond science: during the rescattering [13,14] of an electron, driven by a strong infrared laser-pulse, on its parent ion core. We use dipole approximation for the interaction of a single active electron and its parent ion with the classical electromagnetic field in the length gauge. We approximate the interaction between the active electron and the ion core by a simplistic one-dimensional model with a Dirac delta potential. We measure the entanglement with the von Neumann entropy, and we explore the phase-space structure of the rescattering process by computing the Wigner functions [15-17] of the interacting particles.

## 2. Model and method

The Hamilton operator in terms of the electron and ion-core coordinates (denoted by subscripts *e* and *c*) is given by

$$H = \frac{P_e^2}{2m_e} + \frac{P_c^2}{2m_c} - V\,\delta(X_e - X_c) + e\,E(t)\,(X_e - X_c) \qquad (1)$$

Working with the center of mass and relative coordinates makes the Schrödinger equation separable. We consider the electric field as a 3-cycle laser pulse with sine-squared envelope, having a period of 100 a.u. This pulse excites the atom from the following initial state: the relative state (depending on $x = x_e - x_c$) is the single bound state of the delta potential, while the center of mass has a Gaussian wavefunction with its width fitted to the size of a H atom. The strength of the delta potential is also chosen to ensure that its only bound state has the same energy as the ground state of the H atom. The time evolution of the center of mass is simple spreading, but the evolution the relative particle needs numerical simulation. We use a higher order (Numerov extended) Crank-Nicolson scheme.

## 3. Rescattering in phase space: Wigner functions

The Wigner function is a phase space representation for the state of a quantum system [15,16]. In the simplest case of a pure state in one dimension, its definition is the following, in terms of the coordinate wavefunction $\psi(x)$:

$$W(q,p) = \frac{1}{\pi\hbar} \int_{-\infty}^{\infty} \psi^*(q-x)\psi(q+x)\exp\left(2\frac{i}{\hbar}px\right)dx \qquad (2)$$

The Wigner function is frequently called quasi-probability density function, because its role is similar to the classical probability density function defined over the classial phase space. The Wigner function usually has also negative values, and it is very suitable to explore quantum features, like quantum interference.

Figure 1. shows a snapshot of the Wigner function of the relative particle at a time instant in the middle of the process, as a density plot. The horizontal axis is for position and the vertical axis is for momentum, in atomic units. A small wavepacket has already tunneled out towards left and it is just returning to the parent ion core, while the main wavepacket is emerging and accelerating towards right, from the bound state around the origin of the plot. Note the interference fringes between the wavepackets and around the bound state.

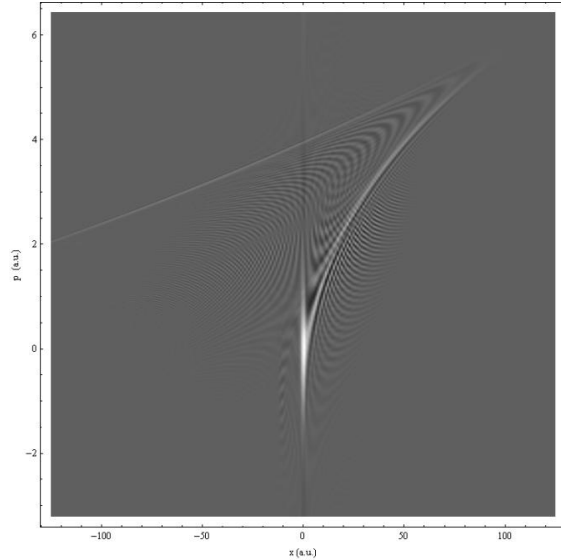

Fig. 1. Density plot of the Wigner function of the relative particle at the time instant of 165 a.u., in the case of CEP = 0. The grey shade filling most of the phase space corresponds to 0, lighter (darker) shades mean positive (negative) function values.

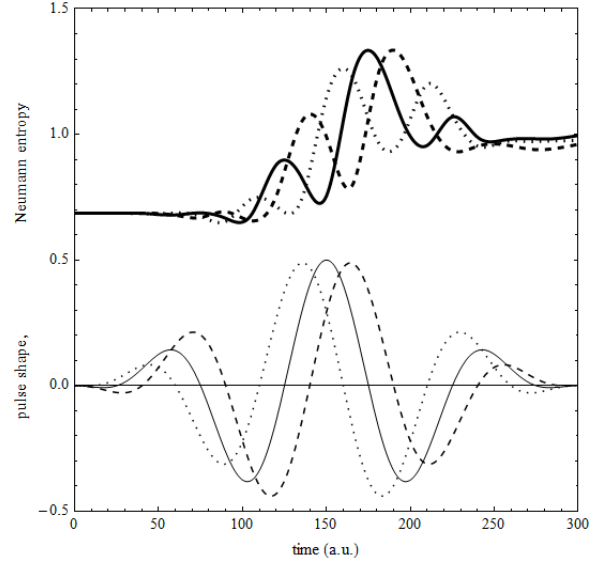

Fig. 2. Neumann entropy versus time (thick lines, top), in comparison with the laser pulse shapes (thin lines, bottom). Solid lines: CEP = 0, dashed lines: CEP = -0.3 $\pi$, dotted lines: CEP = 0.3 $\pi$. See the text below for explanation.

### 4. Quantum entanglement

We use the Neumann entropy to quantify the entanglement of the quantum state of the electron – core system:

$$S(t) = -\mathrm{Tr}[\hat{\rho}_e(t) \ln \hat{\rho}_e(t)] = -\sum_k p_k(t) \ln p_k(t) \qquad (3)$$

Here $p_k(t)$ are the eigenvalues of the electron's reduced one-particle density matrix $\hat{\rho}_e(t)$. Figure 2. shows the time evolution of entanglement during the rescattering process: here we plot the Neumann entropy as the function of time, for 3 different values of the carrier – envelope phase (CEP), given in the figure caption. Note that the local maxima of the oscillations coincide with the zero crossings of the electric field of the corresponding laser pulse.

### Acknowledgments

This research has been granted by the Hungarian Scientific Research Fund OTKA under Contracts No. T81364, and by the „TAMOP-4.2.1/B-09/1/KONV-2010-0005 project: Creating the Center of Excellence at the University of Szeged" supported by the EU and the European Regional Development Fund.